\documentclass[letterpaper]{article} 
\usepackage{aaai2026}  
\usepackage{times}  
\usepackage{helvet}  
\usepackage{courier}  
\usepackage[hyphens]{url}  
\usepackage{graphicx} 
\usepackage{natbib}  
\usepackage{caption} 
\usepackage{algorithm}
\usepackage{algorithmic}
\usepackage{amsmath,amsthm,amssymb}
\usepackage{multirow}
\usepackage{booktabs}
\usepackage{pgfplots}
\usepackage{subfigure}
\usepackage{pifont}
\usepackage{newfloat}
\usepackage{listings}
\DeclareCaptionStyle{ruled}{labelfont=normalfont,labelsep=colon,strut=off} 
\floatstyle{ruled}
\newfloat{listing}{tb}{lst}{}
\floatname{listing}{Listing}
\title{End-to-end Contrastive Language-Speech Pretraining Model For Long-form Spoken Question Answering}

\author{
    Jiliang Hu\textsuperscript{\rm 2}, 
    Zuchao Li\textsuperscript{\rm 1,\footnotemark[1]}, 
    Baoyuan Qi\textsuperscript{\rm 4}, 
    Liu Guoming\textsuperscript{\rm 4}, 
    Ping Wang\textsuperscript{\rm 3}
}

\affiliations{
    \textsuperscript{\rm 1}School of Artificial Intelligence, Wuhan University, Wuhan, China, \\
    \textsuperscript{\rm 2}Key Laboratory of Aerospace Information Security and Trusted Computing, Ministry of Education, \\School of Cyber Science and Engineering, Wuhan University, Wuhan, China, \\
    \textsuperscript{\rm 3}School of Information Management, Wuhan University, Wuhan, China,\\
    \textsuperscript{\rm 4}Xiaomi, Beijing, China. \\
    \{jilianghu, zcli-charlie\}@whu.edu.cn, \{qibaoyuan, liuguoming\}@xiaomi.com, wangping@whu.edu.cn
}

\usepackage{bibentry}

\begin{document}

\maketitle

\begin{abstract}
Significant progress has been made in spoken question answering (SQA) in recent years. However, many existing methods, including large audio language models, struggle with processing long audio. Follow the success of retrieval augmented generation, a speech-related retriever shows promising in help preprocessing long-form speech. But the performance of existing speech-related retrievers is lacking. To address this challenge, we propose CLSR, an end-to-end contrastive language-speech retriever that efficiently extracts question-relevant segments from long audio recordings for downstream SQA task. Unlike conventional speech-text contrastive models, CLSR incorporates an intermediate step that converts acoustic features into text-like representations prior to alignment, thereby more effectively bridging the gap between modalities. Experimental results across four cross-modal retrieval datasets demonstrate that CLSR surpasses both end-to-end speech related retrievers and pipeline approaches combining speech recognition with text retrieval, providing a robust foundation for advancing practical long-form SQA applications.
\end{abstract}

\renewcommand{\thefootnote}{\fnsymbol{footnote}}
\footnotetext[1]{Corresponding author.}
\renewcommand {\thefootnote} {\arabic {footnote}} 

\begin{links}
    \link{Code}{https://github.com/193746/CLSR}
\end{links}



\section{Introduction}

\begin{figure}[htbp]
	\centering 
	\includegraphics[scale=0.4]{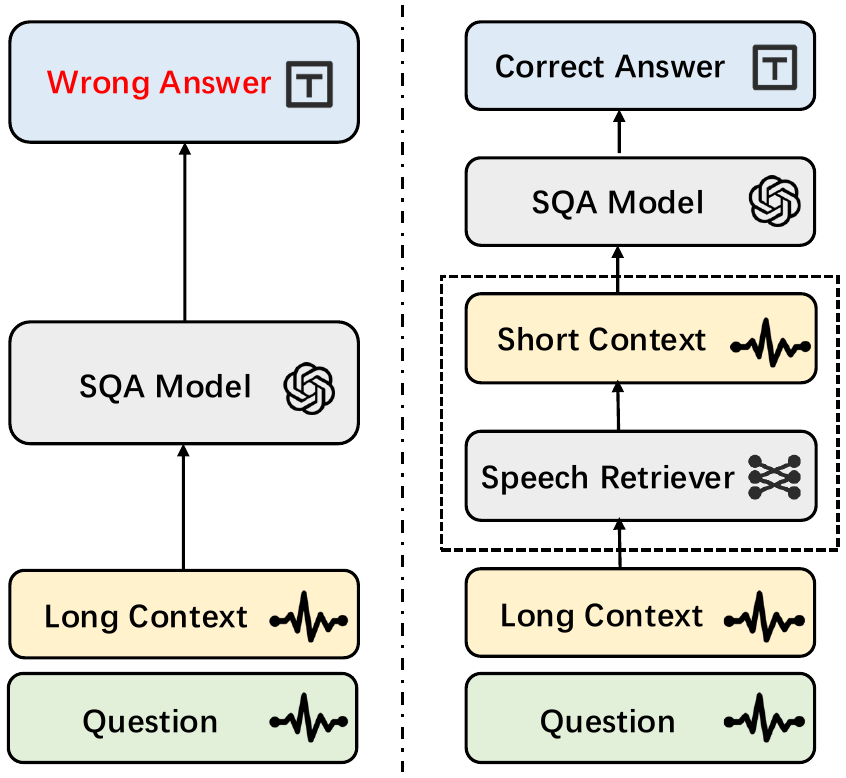} 
	\caption{Using a speech retrieval model to simplify long audio context into several key audio segments can help improve the quality of subsequent LALM response.} 
    \label{intro}
\end{figure}

The question answering (QA) task requires the model to find the answer to a question from a given context. When the answer is a specific segment within the context, the task is classified as extractive QA; conversely, if the answer cannot be directly derived from the context and necessitates additional reasoning by the model, it is termed abstractive QA \citep{shih2023gsqa}. In the realm of spoken question answering, the context is presented in audio format \citep{li2018spoken}, and certain complex SQA tasks also require the questions to be delivered in audio format \citep{shon2022slue}. Despite advancements in SQA methodologies \citep{lee2019mitigating,you2022end}, the majority of existing SQA models are limited to processing short audio segments (under one minute). However, many real-world dialogue scenarios, such as meetings, lectures, and online discussions, often involve voice recordings that exceed ten minutes, posing challenges for current SQA techniques.


The development of large language models (LLMs) is advancing rapidly. Notable examples, such as GPT \citep{brown2020language} and LLaMA \citep{touvron2023llama}, among others \citep{li2023batgpt,yao2024sirllm}, have demonstrated significant success across various traditional natural language processing (NLP) tasks, including question answering. In the speech domain, numerous large audio language models (LALMs) have emerged, showcasing remarkable capabilities in speech comprehension \citep{chu2023qwen,radford2023robust}. The retrieval-augmented generation (RAG) paradigm \citep{lewis2020retrieval} enhances LLMs' natural language understanding by integrating external knowledge \citep{gupta2024comprehensive}. Specifically, RAG employs a retriever to assess the similarity between user queries and segments within a knowledge database, selecting the top-k most relevant segments as supplementary context for the LLM. This approach enables the LLM to better comprehend queries and generate more accurate responses. Currently, RAG is commonly used for long-context reasoning tasks \citep{li2025dialogue,guo2025tom}. For example, in question-answering tasks that involve lengthy input contexts like full articles, RAG works by identifying and retrieving the most pertinent context segments. This process minimizes the inclusion of irrelevant information, which can otherwise introduce errors into the answer and reduce inference speed. Given the effectiveness of RAG in text-based long-context QA, a pertinent question arises for long-form SQA: Can RAG be similarly employed to extract problem-relevant segments from audio inputs to serve as context for subsequent LALM processing?


In this paper, we propose CLSR, an end-to-end (E2E) contrastive language-speech retriever designed to distill lengthy speech recordings into a selection of audio clips that are most pertinent to a given query. These audio clips are used for subsequent LALM inference. Unlike conventional E2E speech-text contrastive learning models, CLSR does not endeavor to align acoustic representations and text representations directly. Instead, it first converts acoustic representations into text-like representations, which are then aligned with actual text representations. The extraction of text-like representations primarily employs continuous integrate-and-fire (CIF) to map acoustic representations from temporal steps to token numbers, followed by the application of a vector quantizer (VQ) based adaptor to refine these acoustic representations into text-like forms. We conduct a comparative analysis of CLSR against standard E2E speech-text retrievers and pipeline retrievers, which integrate speech-to-text models with text contrastive learning models, across four datasets: Spoken-SQuAD, LibriSQA, SLUE-SQA-5, and DRCD. The experimental findings demonstrate that CLSR exhibits superior retrieval performance, suggesting that the use of text-like representations as an intermediary between acoustic and text representations enables CLSR to more effectively discern the similarities and distinctions between these two modalities, thereby enhancing the accuracy of pairing speech with text or speech with speech. The contributions of this paper are as follows:
\begin{enumerate}
\item[(1)] To our knowledge, this study represents the inaugural introduction of the RAG concept within the domain of SQA and its application to address challenges associated with lengthy speech inputs. 
\item[(2)] The proposed model initially transforms acoustic representations into text-like representations, subsequently aligning these text-like representations with text representations, which effectively mitigates modal discrepancies and facilitates cross-modal alignment.
\item[(3)] The CLSR demonstrates superior performance compared to both E2E and cascade speech retrieval systems on four datasets: Spoken-SQuAD, LibriSQA, SLUE-SQA-5, and DRCD.
\end{enumerate}


\section{Related Work}
Currently, there are many works related to SQA. \citet{chuang2019speechbert} propose a pre-trained model called SpeechBERT for the E2E SQA task. Through the training stage called initial phonetic spatial joint embedding for audio words, it aligns the generated audio embeddings with the text embeddings produced by BERT \citep{devlin2019bert} within the same hidden space. \citet{ shih2023gsqa} introduce GSQA, which enables the SQA system to perform abstractive reasoning. They first utilize HuBERT \citep{hsu2021hubert} to convert the input speech into discrete units, and then employ a sequence-to-sequence SQA model finetuned from the text QA model LongT5 \citep{guo2022longt5} to generate answers in the form of discrete units. \citet{lin2024speechdpr} focus on open-domain SQA in scenarios whose paired speech-text data is unavailable. They propose SpeechDPR, which utilizes a bi-encoder retriever framework and learns a sentence-level semantic representation space by extracting knowledge from a combined model of automatic speech recognition (ASR) and text retrieval. \citet{johnson2024efficient} introduce a retriever that employs deep Q-learning to bypass irrelevant audio segments in longer audio files, thereby enhancing the efficiency of SQA. The latter two articles are related to retriever, which is similar to our paper; however, they have limitations: the performance of the former is inferior to that of the pipeline model, and the latter poorly adapts to the high-dimensional, complex feature space of audio, easily falls into local optima during training due to unbalanced exploration and exploitation.


Since the inception of GPT, RAG has advanced rapidly \citep{zhao2025dac}, while research on speech RAG has been comparatively limited. \citet{yang2024srag} utilize RAG for spoken language understanding (SLU). They first employ a pre-trained ASR encoder to extract acoustic features. Next, they perform a similarity calculation to identify audio-text label pairs in the training set that are similar, subsequently incorporating this label information into the SLU decoder through a cross-attention mechanism. \citet{wang2024retrieval} propose a joint speech and language model based on RAG, which enhances performance on the named entity recognition task. They compute the similarity between the input speech query embeddings and the entity embeddings in the database to extract the K entities most relevant to the query, using these entities as additional inputs to the model. Currently, there is no SRAG model designed for long-form SQA task.

\section{Method}
\subsection{Preliminary}
\begin{figure}[htbp]
	\centering 
	\includegraphics[scale=0.37]{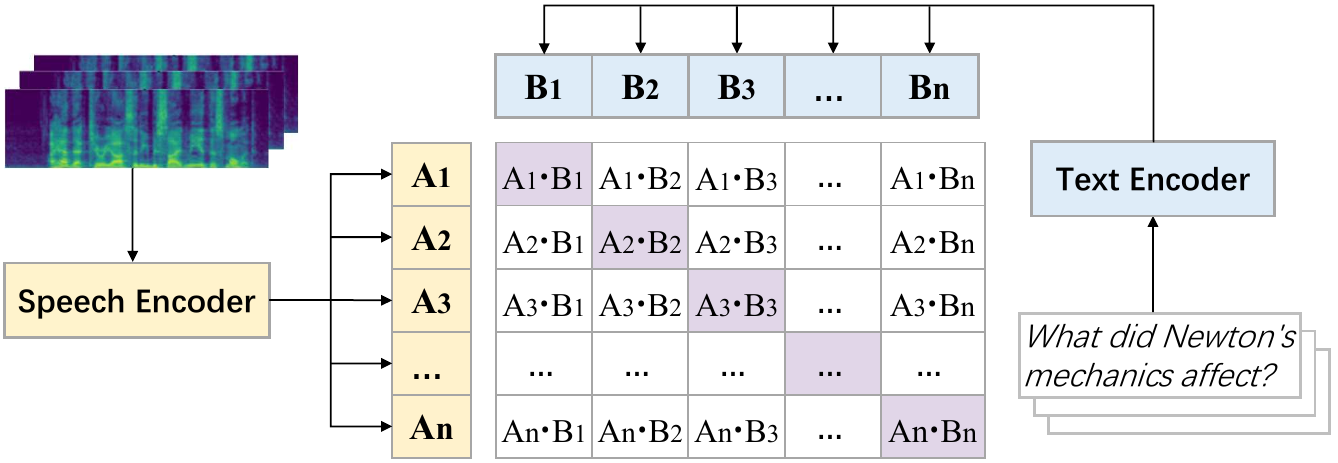} 
	\caption{The architecture of typical E2E speech-text contrastive model.} 
    \label{baseline}
\end{figure}

Consider the SQA task, where questions are presented in text format and contexts are delivered in speech format. Figure \ref{baseline} illustrates the architecture of a typical E2E audio-text contrastive model, such as CLAP \citep{wu2023large}. This model employs a speech encoder and a text encoder to derive sample-level global features, respectively. Let \( \mathcal{A} = \{A_1, A_2, \dots, A_n\} \) denote the set of acoustic feature vectors extracted from a batch of speech contexts. Similarly, let \( \mathcal{B} = \{B_1, B_2, \dots, B_n\} \) denote the set of textual feature vectors extracted from a batch of question texts. The pairwise similarity \( S_{i,j} \) between the speech feature \( A_i \) and the text feature \( B_j \) is quantified using cosine similarity, formulated as follows where \( \|\cdot\| \) denotes the L2 norm:

$$
S_{i,j} = \frac{A_i \cdot B_j}{\|A_i\| \cdot \|B_j\|}
$$


Contrastive learning models of this type focus on deriving sample-level global features for each speech and text sample. Two common methods exist for this purpose. The first introduces a trainable {\tt CLS} token during encoding; the representation corresponding to this token is then used as the global sample feature. The second method computes the average of all frame/token-level features to produce a single global feature vector.

During training, the model optimizes the symmetric contrastive loss (InfoNCE), which aims to maximize the similarity of matching positive pairs (\( A_i, B_i \)) while minimizing the similarity of mismatched negative pairs within the batch. This loss function consists of two symmetric terms: one for text-to-speech retrieval and the other for speech-to-text retrieval. The final formulation is as follows:

$$
\mathcal{L}_{CL} = -\frac{1}{2n} \sum_{i=1}^n \left( \log \frac{e^{S_{i,i}}}{\sum_{j=1}^n e^{S_{i,j}}} + \log \frac{e^{S_{i,i}}}{\sum_{j=1}^n e^{S_{j,i}}} \right)
$$


\subsection{Overview}
\begin{figure}[htbp]
	\centering 
	\includegraphics[scale=0.51]{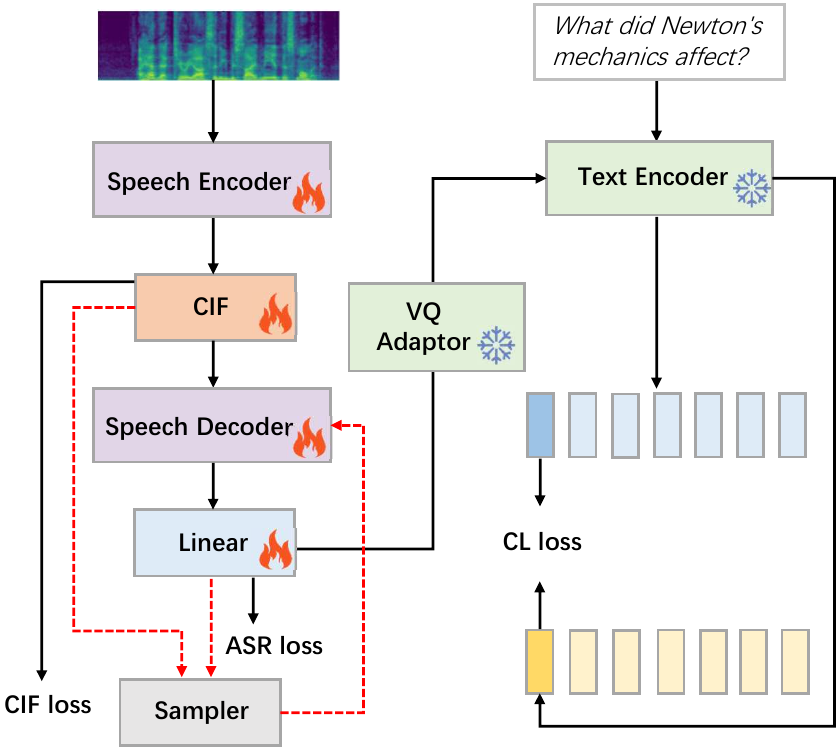} 
	\caption{The architecture of proposed model, CLSR. The red line is only used during training.} 
    \label{clsr}
\end{figure}


Let $X =\{x_1, x_2, \dots, x_t\}$ denote the speech context, represented as a sequence of $t$ frames. Let $Z = \{z_1, z_2, \dots, z_m\}$ denote the question text, represented as a sequence of $m$ tokens, where each token $z_i$ belongs to a vocabulary $V$ (i.e., $z_i \in V$). Figure \ref{clsr} shows the specific architecture of CLSR. The left half features a non-autoregressive attention encoder-decoder (AED) framework based on CIF \citep{dong2020cif}. It takes the speech context $X$ as input and produces the corresponding token probability distribution $D$, $D=\{d_1,d_2,d_3,…,d_n\}$. Both the speech encoder and decoder adopt the SAN-M \citep{gao2020san} structure, which is a specialized Transformer \citep{vaswani2017attention} layer that integrates a self-attention mechanism with deep feed-forward sequential memory networks (DFSMN). Initially, the framework uses the speech encoder to extract acoustic features $H^s$.
$$
H^s=SEncoder(X)
$$

Next, it maps $H^s$ from the time step to the number of tokens through the soft and monotonic alignment mechanism, CIF, obtaining an acoustic representation $E^a$, which is aligned with the token probability distribution.
$$
E^a=CIF(H^s)
$$

Then, as shown in the following formula, it predicts the corresponding token distribution through the speech decoder and a fully connected layer. In addition, following \citet{gao2022paraformer}, we use a sampler to optimize the training process of this framework. The sampler does not contain any learnable parameters and is designed to enhance the context modeling capability of the decoder by sampling text features into $E^a$. It will be elaborated on in subsequent chapter.
$$
D=W\cdot Decoder(H^s,E^a)+b
$$

Next, we utilize the VQ adaptor to map the token distribution to text-like embeddings.

$$
E^{Y^{'}}=VQAdaptor(D)
$$

The right half of CLSR is a Transformer-based text encoder that receives either text embeddings or text-like embeddings as input and output corresponding text representations. We obtain the sentence-level representation by inserting the {\tt CLS} token. When aligning the text question with the speech context, we input the text-like embeddings $E^{Y^{'}}$ of the context and the text embeddings $E^Z$ of the question into the text encoder to obtain their respective sentence-level representations. We then use cosine similarity to evaluate the similarity between them.


$$
S_{Y^{'},Z}=\frac{TEncoder(E^{Y^{'}}) \cdot TEncoder(E^Z)}{||TEncoder(E^{Y^{'}})||\cdot||TEncoder(E^Z)||}
$$

\subsection{Continuous Integrate-and-Fire}
\begin{figure}[htbp]
	\centering 
	\includegraphics[scale=0.6]{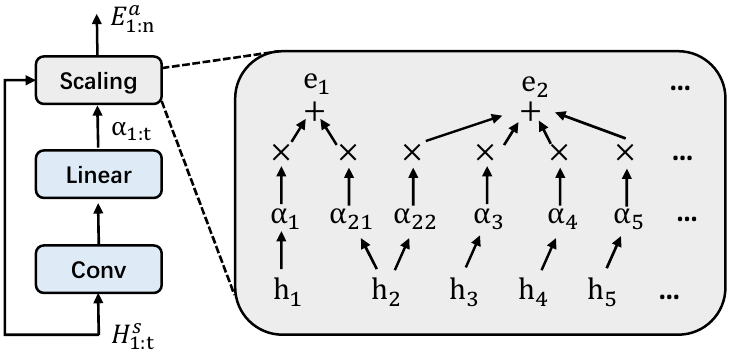} 
	\caption{The explanation of CIF workflow. The gray box on the right shows an example of CIF, where $\alpha=\{0.8,0.3,0.4,0.4,0.1\}$ and the threshold $\beta$=1.}
    \label{cif}
\end{figure}

Figure \ref{cif} explains the workflow of the CIF. Through convolution operations and linear mapping, it calculates the weight distribution $\alpha$, $\alpha=\{\alpha_1,\alpha_2,\alpha_3,…,\alpha_t \mid \alpha_i\in [0,1]\}$. Each $\alpha_i$ shows the valid information contained in relevant $h_i$ of the acoustic feature $H_{1:t}^{s}$.

$$
\alpha_{1:t}=W \cdot conv(H_{1:t}^{s})+b
$$

Then, it gathers the weights and combines $H_{1:t}^{s}$ until the total weight hits a specified threshold $\beta$, signaling that an acoustic boundary has been attained. When reaching the threshold, if the current state of $\alpha$ overflows, it will be used for the next round of weight accumulation.




\subsection{Sampler}
To enhance the capacity of the selected non autoregressive AED framework to model token probability distributions, we introduce a training optimization module called the sampler. When the sampler is enabled, the training process of the framework consists of two rounds. In the first round, we do not utilize the sampler; instead, we directly employ the acoustic features $E^a$ obtained from the CIF module to predict the probability distribution of tokens. By applying $argmax$, we can derive the transcription result $Y^{asr}$.
$$
Y^{asr}=\underset{y_i\in V}{\arg\max}(W\cdot Decoder(H^s,E^a)+b)
$$
Next, we proceed to the second round of training and initiate sampling. We first compare the ASR output $Y^{asr}$ with the ground-truth context $Y^{con}$ to identify tokens containing transcription errors and their respective positions. Then merge the correct embeddings of erroneous tokens from $E^c$ (the embeddings of $Y^{con}$) into the acoustic features $E^a$ through selective replacement, generating semantic features $E^s$. This process is formalized as: 

$$
E^s[:\ ,i,:\ ] = 
\begin{cases} 
E^c[:\ ,i,:\ ]& \text{if token } i \text{ is erroneous} \\
E^a[:\ ,i,:\ ] & \text{otherwise}
\end{cases}
$$

After generating $E^s$, we complete the sampling step. It is important to note that not every correct embedding from an erroneous token will be incorporated into $E^a$; this decision is governed by the mixing ratio $\lambda$ ($\lambda \in (0,1)$).

$$
E^s=Sampler(E^a,E^c,\lceil \lambda \sum_{i=1}^{N}(y_{i}^{asr} \neq y_{i}^{con}) \rceil)
$$

Afterwards, use $E^s$ instead of $E^a$ to calculate the probability distribution of the tokens.
$$
D^{'}=W\cdot Decoder(H^s,E^s)+b
$$

It is important to note that during the initial training phase, no gradient backpropagation is performed, and $Y^{asr}$ is solely utilized to determine the sampling number for the sampler. $D^{'}$ obtained in the second phase is then used to calculate the ASR loss.

Regarding the real text embeddings $E^c$, \citet{gao2022paraformer} uses the embedding layer of the speech decoder to derive them. However, in our proposed model, this layer is not trained, and its weights may not effectively represent the text embedding space. Consequently, we use the weights of the linear layer, which is used to generate the probability distribution of the tokens, to calculate $E^c$.

\subsection{Adaptor}
\begin{figure}[htbp]
	\centering 
	\includegraphics[scale=0.8]{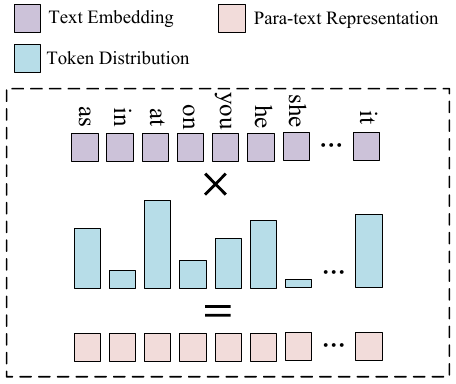} 
	\caption{The mapping process of the adaptor.} 
    \label{adaptor}
\end{figure}







After obtaining the probability distribution $D$ of the tokens, we use an adaptor to map it to the text-like embedding $E^{Y'}$. The adaptation involves two steps: quantization and mapping. The quantization converts each token's probability distribution into the index of the highest-probability token in vocabulary $V$. Following \citet{shih2023speechclip}, we first select the token index $q_i$ with the highest probability from each distribution $d_i$:

$$
q_i = \underset{j \in [1, |V|]}{\arg\max} \, d_{i,j}
$$

Since $q_i$ is non-differentiable, directly using it would break the computational graph. For gradient propagation, we use the temperature-scaled softmax distribution $\tilde{p}_i$, where $\gamma=0.1$ is a hyper-parameter:

$$
\tilde{p}_i = softmax(d_i / \gamma)
$$

Through the straight-through gradient estimator \citep{bengio2013estimating}, we combine the discrete token $q_i$ (represented as a one-hot vector) with the continuous approximation $\tilde{p}_i$ while maintaining gradient flow. The formula is as follows, where $sg(x)=x$ and $\frac{d}{dx}\mathrm{sg}(x)=0$ is the stop-gradient operator:

$$
q_i^{\text{st}} = q_i + \tilde{p}_i - \text{sg}(\tilde{p}_i)
$$

The quantized token representation $q_i^{\text{st}}$ are collected into a matrix $Q^{st}$. Next, we map $Q^{st}$ to the embedding layer of the text encoder. The specific operation, illustrated in Figure \ref{adaptor}, involves matrix multiplication with the embedding layer weights $W^{te}$:

$$
E^{Y^{'}}=Matmul(Q^{st},W^{te})
$$

\subsection{Loss Function}
The adopted framework calculates three loss functions during training: cross-entropy (CE), mean absolute error (MAE), and minimum word error rate (MWER) loss. CE and MWER are used to optimize the model's transcription ability, while MAE facilitates convergence of the CIF. According to \citet{gao2022paraformer}, the loss function for the ASR part is:
$$
\mathcal{L}_{ASR}=\gamma\mathcal{L}_{CE}+\mathcal{L}_{werr}^N(x,y^*)
$$
$$
\mathcal{L}_{werr}^N(x,y^*)=\sum_{y_i \in sample} p(y_i \mid x)[\mathcal{W}(y_i,y^*)-W]
$$

We also use InfoNCE loss to optimize the model's ability in aligning the question representation with the context representation. The overall loss function can be expressed as follows, where $\alpha$ and $\beta$ are parameters that control the proportions of the CIF loss and the contrastive loss, with $\alpha \in (0,1)$ and $\beta \in (0,1)$.
$$
\mathcal{L}_{total}=(1-\alpha-\beta)\mathcal{L}_{ASR}+\alpha\mathcal{L}_{CIF}+\beta\mathcal{L}_{CL}
$$

\section{Experiment}
\subsection{Configuration}

\begin{table}[htbp]
    \renewcommand{\arraystretch}{1.3} 
	\small
	\resizebox{\columnwidth}{!}{
		\begin{tabular}{cccccccc}	 
			\hline
			\multirow{2}{*}{Dataset}&\multirow{2}{*}{Language}&\multicolumn{2}{c}{Type}&\multicolumn{3}{c}{Size} \\
            \cmidrule(lr){3-4}\cmidrule(lr){5-7}
            &&Question&Context&Train&Val&Test\\
			\hline
			Spoken-SQuAD&English&Text&Speech&37,107&5,351&-\\
            Spoken-SQuAD*&English&Text&Speech&29,227&3,884&-\\
            LibriSQA&English&Text&Speech&104,014&2620&-\\
            SLUE-SQA-5&English&Speech&Speech&46,186&1,939&2,382\\
            DRCD*&Chinese&Speech&Speech&25,321&1,425&-\\
			\hline		
		\end{tabular}
	}
    \caption{Experimental datasets. Datasets marked with asterisks are filtered to ensure one-to-one correspondence between problems and contexts.}
    \label{datasets}
	\centering

\end{table}

We conduct experiments on four datasets: Spoken-SQuAD \citep{li2018spoken}, LibriSQA \citep {zhao2024librisqa}, SLUE-SQA-5 \citep{shon2022slue}, and DRCD. Table \ref{datasets} displays detailed information about these datasets. \citet{li2018spoken} use Google text-to-speech (TTS) system to generate the spoken versions of the articles in SQuAD \citep{rajpurkar2016squad}. Given that SQuAD is a many-to-one dataset, where multiple questions correspond to the same context, it is unsuitable for training text-speech retrievers. So, we filter the original Spoken-SQuAD dataset to ensure that each question corresponds one-to-one with its context; the filtered dataset is referred to as Spoken SQuAD*. LibriSQA is adapted from the ASR dataset LibriSpeech \citep{panayotov2015librispeech}. The authors input the textual document of each speech segment from LibriSpeech into ChatGPT and request the generation of corresponding text question-answer pairs. We use the first part of LibriSQA, which presents questions without options, and the answers are complete sentences. SLUE-SQA-5 is adapted from five text QA datasets, and both the questions and contexts consist of authentic audio recordings. DRCD \citep{shao2018drcd} is originally a Chinese QA dataset. Similar to SQuAD, it is also a many-to-one dataset. We first filter it into a one-to-one dataset and then use the TTS model \citep{li2020robutrans} to synthesize the speech versions of each question-context pair for its training set. \citet{lee2018odsqa} offer spoken version of DRCD's development set, which we use for testing.


CLSR is built with Paraformer \citep{gao2022paraformer} (220M) and frozen BGE-base \citep{chen2024bge} (109M). Two kinds of baselines are compared: an E2E text‑speech contrastive model like Figure \ref{baseline} and a cascaded model that first transcribes speech with an ASR module followed by a text QA module. For the former, CLAP and SpeechDPR are selected; for the latter, we employ Whisper \citep{radford2023robust} (244M) for ASR and BGE‑base for text QA. Evaluation uses word error rate (WER) for ASR performance and top‑k retrieval recall for retrieval ability. Experiments are conducted on FunASR \citep{gao2023funasr} and ModelScope. The loss weights $\alpha$ and $\beta$ are set to $\frac{1}{3}$. Models are trained to convergence with the Adam optimizer at a learning rate of 5e-5.



\subsection{Main Result}
\begin{table*}[htbp]
    \renewcommand{\arraystretch}{1.1} 
	\small
	\resizebox{\linewidth}{!}{
		\begin{tabular}{cccccccccccc}	 
			\hline
			\multirow{2}{*}{Dataset} &\multirow{2}{*}{Model} &\multirow{2}{*}{Paradigm} &\multicolumn{2}{c}{Type} &ASR &\multicolumn{3}{c}{Q-C Retrieval ($\uparrow$)} & \multicolumn{3}{c}{C-Q Retrieval ($\uparrow$)}\\
			\cmidrule(lr){4-5}\cmidrule(lr){6-6}\cmidrule(lr){7-9}\cmidrule(lr){10-12}
            &&&Question&Context&WER ($\downarrow$)&R@1&R@5&R@10&R@1&R@5&R@10 \\
			\hline
            \multirow{3}{*}{Spoken-SQuAD*}&BGE&E2E&Text&Text&0&67.12&85.20&89.44&65.63&84.14&89.06\\
            &CLAP&E2E&Text&Speech&-&2.93&9.92&14.84&3.20&10.15&15.23\\ 
            &Whisper+BGE&Pipeline&Text&Transcript&19.39&69.93&86.61&90.53&\textbf{67.97}&\textbf{85.76}&89.65\\
            &CLSR&E2E&Text&Speech&\textbf{15.14}&\textbf{70.03}&\textbf{86.90}&\textbf{90.68}&67.84&85.69&\textbf{90.17}\\
            \hline
            \multirow{3}{*}{LibriSQA}&BGE&E2E&Text&Text&0&86.91&94.31&95.92&86.87&94.73&96.60\\
            &CLAP&E2E&Text&Speech&-&0.04&0.19&0.38&0.08&0.19&0.50\\
            &Whisper+BGE&Pipeline&Text&Transcript&4.32&83.70&93.28&94.92&85.15&93.40&95.27\\
            &CLSR&E2E&Text&Speech&\textbf{4.09}&\textbf{85.04}&\textbf{93.36}&\textbf{95.04}&\textbf{85.53}&\textbf{94.01}&\textbf{95.57}\\
            \hline
            \multirow{3}{*}{SLUE-SQA-5}&BGE&E2E&Text&Text&0&38.71&72.26&84.34&35.68&70.11&82.28\\
            &CLAP&E2E&Text&Speech&-&11.17&28.67&38.16&11.21&28.59&38.12\\
            &SpeechDPR&E2E&Speech&Speech&-&-&-&19.94*&-&-&-\\
            &Whisper+BGE&Pipeline&Transcript&Transcript&36.41&29.98&60.41&72.71&29.85&60.75&\textbf{73.47}\\
            &CLSR&E2E&Speech&Speech&\textbf{16.69}&\textbf{30.65}&\textbf{62.19}&\textbf{74.43}&\textbf{29.89}&\textbf{62.18}&73.05\\
            \hline
            \multirow{2}{*}{DRCD*}&BGE&E2E&Text&Text&0&90.67&97.12&98.74&89.26&97.75&98.39\\
            &CLAP&E2E&Text&Speech&-&0.35&1.33&2.95&0.35&1.12&1.61\\
            &CLSR&E2E&Speech&Speech&\textbf{5.56}&76.21&87.79&90.03&75.23&88.21&91.51\\
            \hline
		\end{tabular}
	}

	\centering
    \caption{Main results of the proposed model across four datasets. Results for BGE are included as a reference benchmark, showing theoretical limits under optimal ASR conditions (100\% accuracy). The SpeechDPR's paper only provides the result of R@20. CLAP is composed of HTSAT \citep{chen2022hts} and RoBERTa \citep{liu2019roberta}.}
    \label{mainresult}
\end{table*}

Table \ref{mainresult} shows the comparison results of CLSR and other models across four datasets. We additionally provide the results of using BGE for clean text question-context retrieval. In terms of E2E text-to-speech contrastive models, the results of CLSR are significantly better than those of CLAP and SpeechDPR. We found that CLAP cannot learn the relevance between text questions and speech contexts effectively on four datasets,  suggesting that CLAP is not well-suited for text-to-speech content alignment. In fact, CLAP is more appropriate for sound and text alignment.  

SpeechDPR is committed to using text-less data for training. Although they use ASR models and text QA models for knowledge distillation, the scarcity of data hampers its ability to achieve optimal performance. It is worth noting that we do not conduct large-scale pre-training prior to training CLSR. All leading contrastive learning models, such as BGE, have undergone extensive pre-training, which enhances their retrieval capabilities. Nevertheless, CLSR still achieves results that are second only to BGE in clean text retrieval and even surpasses BGE's performance on Spoken-SQuAD*, highlighting the advantages of CLSR's architecture.

Compared to conventional E2E contrastive models that directly perform text-to-speech or speech-to-speech alignment, CLSR utilizes text-like representations to alleviate the differences between speech and text modalities. It first maps speech representations into text-like representations and then aligns these text-like representations with actual text representations (or aligns text-like representations with other text-like representations) within the text modality. Leveraging the robust performance of text contrastive models, this approach enhances the alignment between speech and text (or between speech and speech), thereby facilitating more accurate pairing with the context most relevant to the question.

When conducting a comparative analysis of CLSR and Whisper+BGE, we find that their retrieval performances on three English datasets are quite similar; however, CLSR demonstrates certain advantages. In terms of transcription ability, CLSR significantly outperforms Whisper+BGE. This indicates that the joint training of CLSR effectively optimizes both the ASR module and the contrastive learning module. Given that Whisper's performance in Chinese speech recognition is not exceptional, we have opted not to train Whisper on DRCD*.




\subsection{Ablation Result}

\begin{table*}[htbp]
    \renewcommand{\arraystretch}{1.1} 
	\small
		\begin{tabular}{cccccccccccc}	 
			\hline
			\multicolumn{2}{c}{Pre-train} &\multicolumn{2}{c}{Joint-train} &Post-train &ASR &\multicolumn{3}{c}{Q-C Retrieval ($\uparrow$)} & \multicolumn{3}{c}{C-Q Retrieval ($\uparrow$)} \\
			\cmidrule(lr){1-2}\cmidrule(lr){3-4}\cmidrule(lr){5-5}\cmidrule(lr){6-6}\cmidrule(lr){7-9}\cmidrule(lr){10-12}
            ASR&BGE&VQ&Sampler&BGE&WER ($\downarrow$)&R@1&R@5&R@10&R@1&R@5&R@10\\
			\hline
            \ding{53}&\ding{53}&\ding{53}&\ding{53}&\ding{53}&16.13&15.29&34.14&44.18&15.75&36.11&46.16\\
            \ding{53}&\ding{53}&\ding{51}&\ding{53}&\ding{53}&17.00&42.52&71.46&78.36&46.86&72.66&79.95\\
            \ding{53}&\ding{53}&\ding{51}&\ding{53}&\ding{51}&17.00&45.11&75.31&82.90&48.05&75.82&83.18\\
            \hline
            \ding{53}&\ding{51}&\ding{51}&\ding{53}&\ding{53}&17.00&48.10&78.28&84.98&49.45&76.79&83.42\\
            \ding{53}&\ding{51}&\ding{51}&\ding{53}&\ding{51}&17.00&48.31&78.55&84.73&50.08&77.16&83.68\\
            \hline
            \ding{51}&\ding{51}&\ding{51}&\ding{53}&\ding{53}&16.18&49.00&79.20&85.69&50.31&77.48&84.21\\
            \ding{51}&\ding{51}&\ding{51}&\ding{51}&\ding{53}&15.01&49.65&79.61&85.91&50.59&77.71&84.38\\
            \ding{51}&\ding{51}&\ding{51}&\ding{51}&\ding{51}&\textbf{15.01}&\textbf{49.82}&\textbf{79.63}&\textbf{85.83}&\textbf{50.63}&\textbf{77.69}&\textbf{84.56}\\
            \hline

		\end{tabular}
	\centering
    \caption{Ablation results in Spoken-SQuAD.}
    \label{ablationresult}
\end{table*}

\begin{table*}[htbp]
    \renewcommand{\arraystretch}{1.1} 
	\small
		\begin{tabular}{cccccccccc}	 
			\hline
			\multirow{2}{*}{Dataset} &\multirow{2}{*}{Model} &\multirow{2}{*}{Paradigm} &ASR &\multicolumn{3}{c}{Q-C Retrieval ($\uparrow$)} & \multicolumn{3}{c}{C-Q Retrieval ($\uparrow$)} \\
			\cmidrule(lr){4-4}\cmidrule(lr){5-7}\cmidrule(lr){8-10}
            &&&WER ($\downarrow$)&R@1&R@5&R@10&R@1&R@5&R@10\\
			\hline
            \multirow{2}{*}{Spoken-SQuAD}
            &ParaBGE&E2E&-&17.79&38.68&48.35&17.03&38.31&48.91\\
            &CLSR&E2E&15.01&\textbf{49.82}&\textbf{79.63}&\textbf{85.83}&\textbf{50.63}&\textbf{77.69}&\textbf{84.56}\\
            \hline
            \multirow{2}{*}{LibriSQA}
            &ParaBGE&E2E&-&29.31&50.27&59.70&20.57&39.28&49.28\\
            &CLSR&E2E&4.09&\textbf{85.04}&\textbf{93.36}&\textbf{95.04}&\textbf{85.53}&\textbf{94.01}&\textbf{95.57}\\
            \hline
            \multirow{2}{*}{SLUE-SQA-5}
            &ParaBGE&E2E&-&7.31&21.83&32.75&7.52&21.96&33.12\\
            &CLSR&E2E&16.69&\textbf{30.65}&\textbf{62.19}&\textbf{74.43}&\textbf{29.89}&\textbf{62.18}&\textbf{73.05}\\
            \hline

		\end{tabular}
	\centering
    \caption{Comparison results between traditional E2E contrastive model and CLSR.}
    \label{compareresult}
\end{table*}

To demonstrate the effectiveness of the quantizer and sampler in CLSR, as well as the potential for multi-stage training to improve model performance. We conduct a series of ablation experiments on Spoken-SQuAD. The results are shown in Table \ref{ablationresult}. The first two rows of the results show the value of the quantizer. When the quantizer is not utilized, the model may achieve a lower WER; however, its comparative learning ability significantly diminishes. The top-10 retrieval recall rate of ``CLSR w/o VQ" is only comparable to the top-1 retrieval recall rate of ``CLSR w/ VQ". The results in the sixth and seventh rows show the effectiveness of the sampler. After introducing the sampler, CLSR not only improves retrieval ability, but also improves ASR performance.

Before joint training, we can pre-train the ASR module and the BGE module of CLSR separately. In the experiment, we use 460 hours of clean LibriSpeech data to pre-train Paraformer, and use Spoken-SQuAD's clean text question-context pairs to train BGE. Comparing the second and fourth rows of the experimental results, it is not difficult to find that pre-training the BGE is beneficial; incorporating the pre-trained BGE during joint training enhances various retrieval metrics of the CLSR. In addition, through the comparison between the fourth and sixth rows, it can be found that pre-training Paraformer can improve the model's transcription performance while also slightly improving its retrieval ability. It should be noted that in order to improve the training speed of the model, we froze BGE, which has strong retrieval performance, during joint training. Therefore, we can freeze the ASR module after joint training and train BGE for a few epochs separately, which is called post-train in the table. It is hoped that this approach can make BGE better adapt to the text-like representation provided by the ASR module. Unfortunately, post-train can only slightly improve the performance of the model, as evidenced by rows 2 and 3, 4 and 5, 7 and 8 in the table. In short, through ablation experiments, we have shown that both quantizers and samplers are inseparable for CLSR, and that pre-training the ASR module and BGE module of CLSR is of significant importance.

\begin{figure}[htbp]
	\centering 	
    \begin{tikzpicture}[scale=0.7]	
        \pgfplotsset{every axis legend/.append style={ at={(1.02,1)},
            anchor=north west}}
        \begin{axis}[
            legend style={at={(0.25,0.80)},anchor=north},
            ytick={40,50,...,80},
            height=2.5in, 
            width=4.5in,
            ylabel style={xshift=0mm,yshift=-4mm},
            xlabel style={xshift=0mm,yshift=0mm},
            xlabel={WER},
            ylabel={Recall}] 
            
            \addplot[color=red,mark=*] coordinates{
                (14.75,49.22)
                (14.87,48.87)
                (14.91,48.80)
                (15.10,48.78)
                (16.04,48.59)
                (16.41,48.37)
                (16.94,42.85)
                (17.13,42.25)
                (17.23,42.81)
                (17.63,42.72)
            };
            \addlegendentry{Q-C R@1}
            
            \addplot[color=blue,mark=star] coordinates{
                (14.75,79.69)
                (14.87,79.54)
                (14.91,79.76)
                (15.10,79.57)
                (16.04,79.37)
                (16.41,79.72)
                (16.94,71.58)
                (17.13,71.46)
                (17.23,71.78)
                (17.63,71.59)
            };
            \addlegendentry{Q-C R@5}
            
            \addplot[color=darkgray,mark=square*] coordinates{
                (14.75,50.48)
                (14.87,50.79)
                (14.91,50.31)
                (15.10,50.61)
                (16.04,49.88)
                (16.41,50.18)
                (16.94,47.41)
                (17.13,47.08)
                (17.23,47.19)
                (17.63,42.17)
            };
            \addlegendentry{C-Q R@1}

            \addplot[color=olive,mark=x] coordinates{
                (14.75,77.67)
                (14.87,77.33)
                (14.91,77.56)
                (15.10,77.72)
                (16.04,77.16)
                (16.41,77.41)
                (16.94,73.28)
                (17.13,73.00)
                (17.23,73.30)
                (17.63,72.75)
            };
            \addlegendentry{C-Q R@5}
            
        \end{axis}
    \end{tikzpicture}
	\caption{The correlation between the retrieval ability and speech recognition ability of CLSR.}	
    \label{trendresult}
\end{figure}
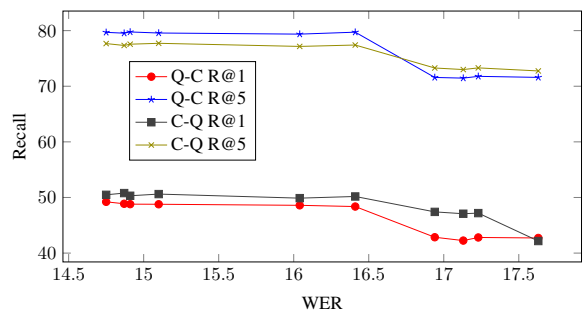


To assess transcription error impact on CLSR's retrieval ability, we evaluate it on Spoken-SQuAD (results in Figure \ref{trendresult}). Lower WER correlates with higher recall rates. Crucially, a WER of ~16.75\% marks a threshold: recall drops significantly above this value.


In order to further demonstrate the superiority of the proposed model over the traditional E2E speech-related contrastive model, which consists of two encoders, we construct a new baseline: ParaBGE, to compare the retrieval capability with CLSR. ParaBGE is composed of the speech encoder from Paraformer and the text encoder from BGE. The sizes of each module in both models are identical to those in CLSR. The experimental results are shown in Table \ref{compareresult}. All retrieval metrics of CLSR far exceed ParaBGE, indicating that CLSR has a stronger question-context alignment ability. Although ParaBGE can optimize parameters towards the direction of aligning question and context representation during training, its performance is not ideal. As we mentioned earlier, such model heavily rely on pre-training with large-scale corpora. However, high-quality speech-text pairs are already very scarce, so for E2E speech related retrieval models, it is difficult to achieve excellent results. However, CLSR alleviates the modal differences between speech and text by using text-like representation as a bridge, shifting the alignment of speech to text alignment. With the powerful generalization ability of text contrastive learning models, it can achieve excellent retrieval capabilities comparable to cascade models and text contrastive models without the need for long-term, large-scale pre-training.

\subsection{Long-form SQA Evaluation}

\begin{table}[htbp]
    \renewcommand{\arraystretch}{1.1} 
	\small
	\resizebox{\columnwidth}{!}{
		\begin{tabular}{ccccc}	 
			\hline
			w/ CLSR & EM ($\uparrow$) &F1 ($\uparrow$) &Cost time (s)&SpeedUP \\
			\hline
            \ding{53}&18.00&23.55&7935.00&1.00X\\
            \ding{51}&\textbf{27.60}&\textbf{35.10}&\textbf{783.00}&\textbf{10.13X}\\
            \hline
		\end{tabular}
	}
	\centering
    \caption{The effectiveness of CLSR when applied to long audio SQA.}
    \label{real SRAG}
\end{table}
To validate our model for real-world long audio SQA—enabling downstream LALMs to simplify inputs and enhance inference speed and accuracy—we conduct SQA tasks on a modified SLUE-SQA-5 dataset using CLSR and Qwen-Audio. To simulate long-context inference, we replace the contextual audio in 500 randomly selected test instances with full documents from the Spoken Wikipedia corpus \citep{kohn2016mining}, each averaging ~30 minutes in length.




Without CLSR, Qwen-Audio directly generates answers from the input speech document and text question. With CLSR, we first segment the speech document into 40-second intervals, assess each segment’s similarity to the text question, and select the most relevant one as Qwen-Audio’s contextual input. Prior to testing, both Qwen-Audio and CLSR were trained using the original training subset of SLUE-SQA-5. The results of the testing are presented in Table \ref{real SRAG}. Following the application of CLSR for long audio reduction, there is a notable enhancement in Qwen-Audio's exact match (EM) and macro-F1 scores for SQA, alongside a tenfold reduction in inference time. These findings underscore the significance of CLSR in the preprocessing of long audio, demonstrating its capacity to not only enhance the inference accuracy of downstream LALM but also to substantially decrease inference time.

\section{Conclusion}

In this paper, we introduce CLSR, an E2E contrastive language-speech retriever designed to distill lengthy speech recordings into a limited number of clips that are most pertinent to a given query. By employing a text-like representation as an intermediary state, CLSR exhibits strong capability of cross-modal question-context alignment. Experimental findings demonstrate that CLSR's retrieval performance significantly outstrips that of existing E2E speech-related retrievers and is competitive with both cascaded models and text-based retrievers.





\section*{Acknowledgements}

This work was supported by the National Natural Science Foundation of
China (No. 62306216), the Fundamental Research Funds for the Central
Universities (No.2042025kf0026), the Technology Innovation Program of Hubei Province (Grant No. 2024BAB043).

\bibliography{custom}

\clearpage
\appendix
\section{Appendix}
\subsection{Does CLSR have an advantage over Whisper+BGE pipeline during runtime?}
CLSR functions as an E2E model, providing a more rapid inference speed in comparison to pipeline models. The speed is a critical runtime metric for a RAG retriever, which must efficiently interface with downstream LALM for long audio inference. We assess the inference speed of CLSR relative to Whisper+BGE across three datasets, with the results detailed in Table \ref{runtime comparison}. While CLSR demonstrates a slight enhancement in transcription and retrieval performance compared to Whisper+BGE, it significantly outperforms in terms of inference speed, indicating that CLSR is more suitable as a RAG retriever.

\begin{table*}[htbp]
    \renewcommand{\arraystretch}{1.1} 
	\small
		\begin{tabular}{ccccccc}	 
			\hline
			Dataset &Model &WER ($\downarrow$) &Q-C R@1 ($\uparrow$)&C-Q R@1 ($\uparrow$) & Cost Time (s) & SpeedUp \\
			\hline
            \multirow{2}{*}{Spoken-SQuAD}&Whisper+BGE&19.39&69.93&\textbf{67.97}&3733.00&1.00X\\
            &CLSR&\textbf{15.14}&\textbf{70.03}&67.84&\textbf{355.00}&\textbf{10.52X}\\
            \hline
            \multirow{2}{*}{LibriSQA}&Whisper+BGE&4.32&83.70&85.15&1470.00&1.00X\\
            &CLSR&\textbf{4.09}&\textbf{85.04}&\textbf{85.53}&\textbf{186.00}&\textbf{7.91X}\\
            \hline
            \multirow{2}{*}{SLUE-SQA-5}&Whisper+BGE&36.41&29.98&29.85&6141.00&1.00X\\
            &CLSR&\textbf{16.69}&\textbf{30.65}&\textbf{29.89}&\textbf{745.00}&\textbf{8.25X}\\
            \hline
		\end{tabular}
	\centering
    \caption{Runtime comparison results between CLSR and Whisper+BGE pipeline.}
    \label{runtime comparison}
\end{table*}

\begin{table*}[htbp]
    \renewcommand{\arraystretch}{1.1} 
	\small
		\begin{tabular}{ccccccccc}	 
			\hline
			\multirow{2}{*}{Dataset} &\multirow{2}{*}{Model} &ASR &\multicolumn{3}{c}{Q-C Retrieval ($\uparrow$)} & \multicolumn{3}{c}{C-Q Retrieval ($\uparrow$)} \\
			\cmidrule(lr){3-3}\cmidrule(lr){4-6}\cmidrule(lr){7-9}
            &&WER ($\downarrow$)&R@1&R@5&R@10&R@1&R@5&R@10\\
			\hline
            \multirow{3}{*}{Spoken-SQuAD}&Whisper+BGE&19.39&69.93&86.61&90.53&\textbf{67.97}&\textbf{85.76}&89.65\\
            &Whisper+BGE*&19.39&69.05&86.10&90.40&67.82&85.48&89.93\\
            &CLSR&\textbf{15.14}&\textbf{70.03}&\textbf{86.90}&\textbf{90.68}&67.84&85.69&\textbf{90.17}\\
            \hline
            \multirow{3}{*}{LibriSQA}&Whisper+BGE&4.32&83.70&93.28&94.92&85.15&93.40&95.27\\
            &Whisper+BGE*&4.32&84.54&92.86&\textbf{95.04}&83.74&93.13&94.85\\
            &CLSR&\textbf{4.09}&\textbf{85.04}&\textbf{93.36}&\textbf{95.04}&\textbf{85.53}&\textbf{94.01}&\textbf{95.57}\\
            \hline
            \multirow{3}{*}{SLUE-SQA-5}&Whisper+BGE*&36.41&29.98&60.41&72.71&29.85&60.75&\textbf{73.47}\\
            &Whisper+BGE*&36.41&23.22&48.49&60.41&23.05&51.34&63.18\\
            &CLSR&\textbf{16.69}&\textbf{30.65}&\textbf{62.19}&\textbf{74.43}&\textbf{29.89}&\textbf{62.18}&73.05\\
            \hline
		\end{tabular}
	\centering
    \caption{Comparison results between E2E and Whisper+BGE pipeline with noisy fine-tuned.}
    \label{noisy comparison}
\end{table*}


\subsection{If the Whisper+BGE pipeline is fine-tuned with noisy Whisper outputs, does CLSR still outperform it?}
To evaluate the performance of the Whisper+BGE pipeline following fine-tuning with noisy data, we transcribe the training data using a trained Whisper model and subsequently provide it to BGE for pre-training. The results of this experimentation are presented in Table \ref{noisy comparison}, where the fine-tuned Whisper+BGE is referred to as Whisper+BGE*. The findings suggest that Whisper+BGE* yields only a marginal improvement in retrieval capability on the LibriSQA dataset, while performance declines on the other two datasets. CLSR continues to demonstrate a superior ability in retrieving speech content.

\subsection{What is the exact mechanism of the sampler?}
Follow \citet{gao2022paraformer}, we use the sampler to optimize the training process of CLSR's ASR module. The sampler combines the correct embeddings of error tokens in $E^c$ into $E^a$, and generates the semantic features $E^s$. The specific pseudocode is shown in Algorithm \ref{sampler mechanism}.
\begin{algorithm*} 
    \renewcommand{\algorithmicrequire}{\textbf{Input:}}
    \renewcommand{\algorithmicensure}{\textbf{Output:}}
	\caption{The mechanism of sampler.} 
	\label{sampler mechanism} 
	\begin{algorithmic}
    \STATE \# $encoder\_out \  [b,t,h1]$ - the output of speech encoder $H^s$
    \STATE \# $ys\_con \ [b,n]$ - real context
    \STATE \# $acoustic\_embeds \ [b,n,h2]$ - acoustic features $E^a$
    \STATE \# $text\_embdeds \ [b,n,h2]$ - text features $E^c$
    \STATE
    \STATE \# Calculate the transcription of the first round training.
    \STATE $decoder\_out = speech\_decoder(encoder\_out,acoustic\_embeds)$
    \STATE $pred\_tokens = decoder\_out.argmax(-1)$
    \STATE \# Calculate the number of correctly decoded tokens.
    \STATE $nonpad\_positions = ys\_con.ne(ignore\_id)$
    \STATE $seq\_lens = (nonpad\_positions).sum(1)$
    \STATE $same\_num = ((pred\_tokens == ys\_con) \& nonpad\_positions).sum(1)$   
    \STATE \# Sample tokens with transcription errors according to sample ratio $\lambda$.
    \STATE $input\_mask = torch.ones\_like(nonpad\_positions)$
    \STATE $bsz, seq\_len = ys\_con.size()$
    \STATE $for \ li\ in\ range(bsz):$
    \STATE    $\quad target\_num = (((seq\_lens[li] - same\_num[li].sum()).float()) * sampling\_ratio).long()$
    \STATE    $\quad if \ target\_num > 0:$
    \STATE    $\quad  \quad input\_mask[li].scatter\_(0,torch.randperm(seq\_lens[li])[:target\_num],0)$
    \STATE \# Merge $E^a$ and $E^c$ into semantic features $E^s$.
    \STATE $input\_mask = input\_mask.eq(1)$
    \STATE $input\_mask = input\_mask.masked\_fill(\sim nonpad\_positions, False)$
    \STATE $input\_mask\_expand\_dim = input\_mask.unsqueeze(2)$
    \STATE $sematic\_embeds = acoustic\_embeds.masked\_fill(\sim input\_mask\_expand\_dim, 0)+$ \\ $text\_embdeds.masked\_fill(input\_mask\_expand\_dim, 0)$
	\end{algorithmic} 
\end{algorithm*}

\subsection{Case Study: Structural Superiority of CLSR over Traditional End-to-End Dual-Encoder Retrievers.}
\begin{figure*}[htbp]
    \centering
    \begin{minipage}{\textwidth}
        \centering
        \includegraphics[scale=0.65]{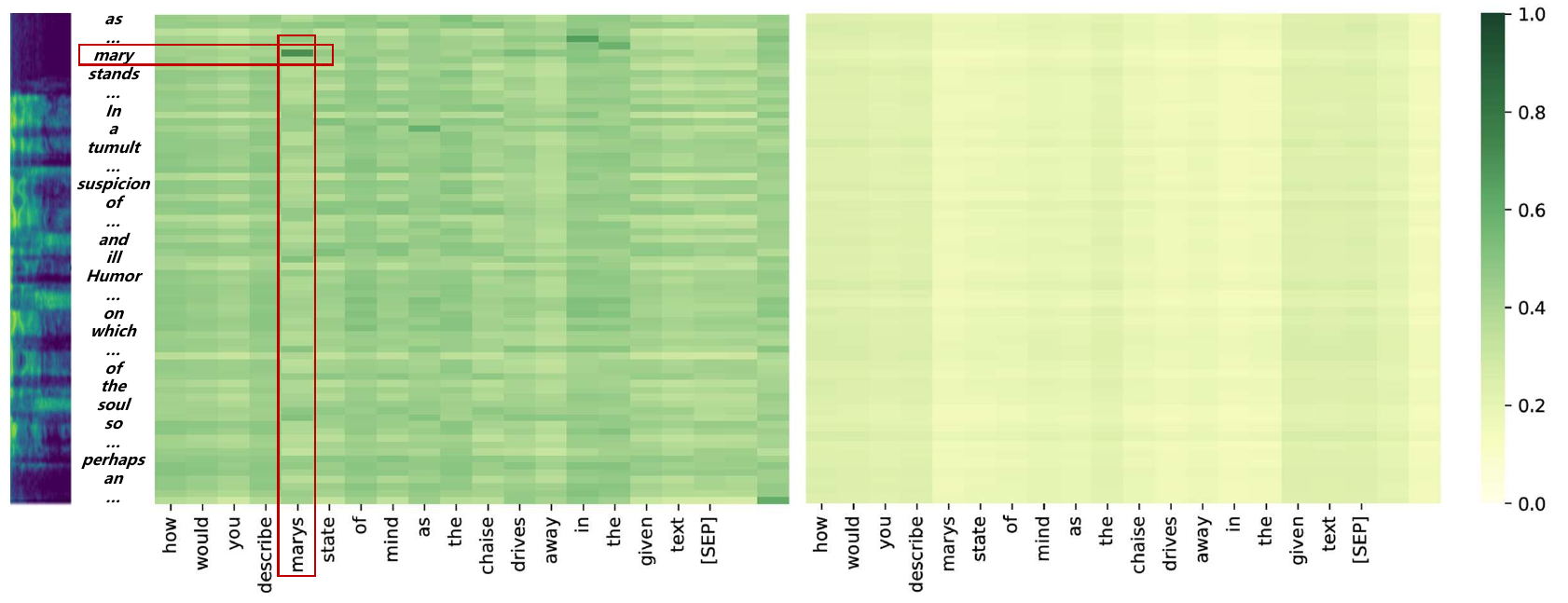} \\[1ex]
        \includegraphics[scale=0.65]{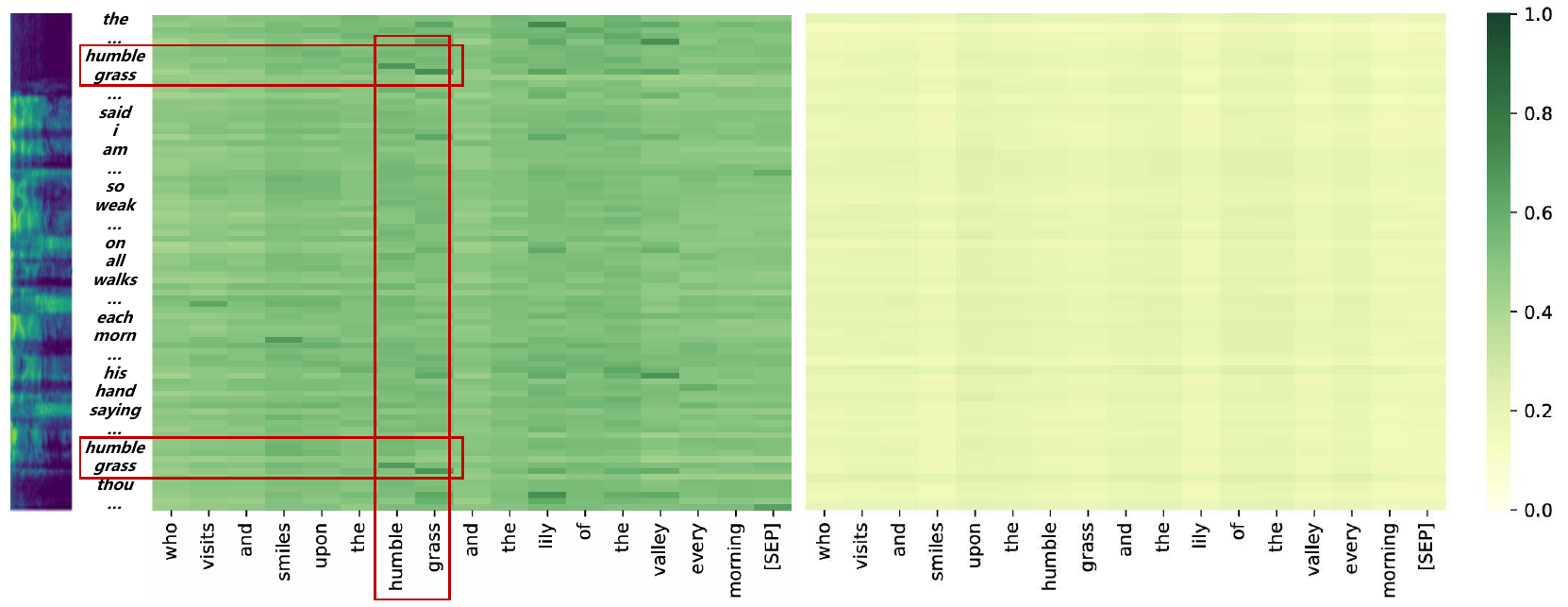}
    \end{minipage}
    \caption{Two comparative case studies between CLSR and ParaBGE. Each case displays two heatmaps with the textual question on the horizontal axis and the speech context's transcription on the vertical axis. The left and right heatmaps correspond to CLSR and ParaBGE, respectively. These heatmaps visualize the cosine similarity scores computed by each model between individual speech frames in the context and word-level tokens in the question, where darker hues indicate higher similarity.}
    \label{casestudy}
\end{figure*}

We select two representative cases from the LibriSQA inference results and visualize the similarity distributions of CLSR and ParaBGE when aligning textual questions with speech contexts in Figure \ref{casestudy}, which demonstrates CLSR's architectural superiority over traditional dual-encoder retrievers. As shown, CLSR achieves precise alignment between semantically related audio-text segments (e.g., matching ``mary" in Case 1 and ``humble grass" in Case 2), whereas ParaBGE exhibits uniformly distributed similarity scores without capturing token-level correspondences. This granular alignment enables CLSR to identify context segments more relevant to queries, enhancing retrieval accuracy. 

CLSR's unique architecture facilitates such granular alignment: It first utilizes the CIF module to project acoustic features from time steps to token positions; then uses VQ-based refinement to convert these features into text-like representations; finally, in text space, leverages pre-trained text retrieval model's power to align text-like representations with ordinary text representations token by token. Since text-like representations retain acoustic-feature similarity, this achieves fine-grained alignment between acoustic and text representations. This architectural superiority is absent in ParaBGE and similar dual-encoder retrievers.

\end{document}